\documentclass{article}
\usepackage{spconf,amsmath,graphicx,bm,amssymb}
\usepackage{algorithm}
\usepackage{color}

\usepackage[pdfstartview=FitH,
CJKbookmarks=true,
bookmarksnumbered=true,
bookmarksopen=true,
colorlinks,
pdfborder=001,
linkcolor=blue,
anchorcolor=blue,
citecolor=blue,
]{hyperref}
\hypersetup{hidelinks}

\usepackage{algorithmic}

\newcommand{\ba}{\mathbf{a}}

\newcommand{\bh}{\mathbf{h}}

\newcommand{\bn}{\mathbf{n}}

\newcommand{\bx}{\mathbf{x}}

\newcommand{\bs}{\mathbf{s}}

\newcommand{\by}{\mathbf{y}}

\newcommand{\bA}{\mathbf{A}}

\newcommand{\bH}{\mathbf{H}}

\newcommand{\RR}{\mathcal{R}}

\newcommand{\T}{{{\mathsf{T}}}}

\newcommand{\I}{\mathcal {I}}

\newtheorem{dingli}{Theorem~}

\def\x{{\mathbf x}}

\title{A novel negative $\ell_1$ penalty approach for multiuser one-bit massive MIMO downlink with PSK signaling}
\name{Zheyu Wu$^{\star,\S}$, Bo Jiang$^{\dag}$, Ya-Feng Liu$^{\S}$, and Yu-Hong Dai$^{\S}$
\thanks{{\color{black}This work was supported in part by the National Natural Science Foundation of China under Grant 12022116 and Grant 12021001.}}
}

\address{$^{\star}$School of Mathematical Sciences, University of Chinese Academy of Sciences, Beijing, China\\[2pt]
$^{\S}$LSEC, ICMSEC, AMSS, Chinese Academy of Sciences, Beijing, China \\[2pt]
    $^{\dag}$School of Mathematical Sciences, Nanjing Normal University, Nanjing, China\\[2pt]
  Email:  \{wuzy, yafliu, dyh\}@lsec.cc.ac.cn, jiangbo@njnu.edu.cn }
  
\begin{document}

\ninept
\maketitle
%
\begin{abstract}
This paper considers the one-bit precoding problem for the multiuser downlink massive multiple-input multiple-output (MIMO) system with phase shift keying (PSK) modulation and focuses on the celebrated constructive interference (CI)-based problem formulation. 
The existence of the discrete one-bit constraint makes the problem generally hard to solve. In this paper, we propose an efficient negative $\ell_1$ penalty approach for finding a high-quality solution of the considered problem.  Specifically, we first propose a novel negative $\ell_1$ penalty model, which penalizes the one-bit constraint into the objective with a negative $\ell_1$-norm term, and show the equivalence between (global and local) solutions of the original problem and the penalty problem when the penalty parameter is sufficiently large.  We further transform the penalty model into an equivalent min-max problem and propose an efficient alternating optimization (AO) algorithm for solving it.  The AO algorithm enjoys low per-iteration complexity and is guaranteed to converge to the stationary point of the min-max problem. Numerical results show that, compared against the state-of-the-art CI-based algorithms, the  proposed algorithm generally achieves better bit-error-rate (BER)  performance with lower computational cost.




%
%
\end{abstract}
\section{Introduction}
Massive multiple-input multiple-output (MIMO), which deploys tens to hundreds of antennas at the base station (BS), is a key technology for significantly improving the spectrum and energy efficiency of 5G and beyond wireless communication systems \cite{massivemimo1}. However, since the number of radio-frequency (RF) chains needs to be scaled up with the number of antennas, the hardware complexity and power consumption would be unaffordably high for practical massive MIMO systems if high-resolution analog-to-digital converters (ADCs)/digital-to-analog converters (DACs) are employed. To deal with such issues, there have been growing interest in the employment of low-resolution ADCs/DACs, especially the cheapest one-bit ones. In particular, the one-bit DAC downlink has attracted a lot of recent research interests \cite{linear1}--\cite{PBB}.

Early works \cite{linear1,linear2,linear3} are based on linear-quantized precoding schemes, in which the one-bit precoders are obtained by directly quantizing the classical linear precoders.  Despite the advantage of their low computational complexities, such linear precoders usually suffer from high symbol error rate floors. As such, there have been emerging works on analyzing and designing nonlinear precoders for one-bit downlink transmission.
In \cite{SQUID,C3PO1,C3PO2}, the authors have focused on the minimum mean square error (MMSE) criterion to formulate the one-bit precoding design problem, and the precoders proposed  therein are shown to greatly enhance the performance of the linear precoders. 
 Recently, the novel idea of constructive interference (CI) \cite{CI1,CItutorial} has been incorporated into one-bit precoding design. There are also some works that directly consider the symbol error probability (SEP) criterion \cite{GEMM, sep2}. In fact, the CI metric is shown to be closely related to the SEP criterion \cite{sep3} and is easier to characterize,  which motivates a new line of research focusing on the CI metric \cite{CIfirst, CImodel, PBB}.
Specifically, in \cite{CIfirst}, the CI-based model for one-bit precoding has been formulated for the first time and a precoder based on linear programming (LP) relaxation named maximum safety margin (MSM) has been developed. Later, the authors in \cite{CImodel} have proposed an alternative CI-based model, known as the symbol-scaling model, which admits a simpler formulation and is shown in \cite{ciequivalent} to be equivalent to the previous model.   

The existing state-of-the-art algorithms \cite{CImodel, PBB} for the CI metric are mainly based on the LP relaxation of the symbol scaling model.  These algorithms generally consist of two stages: in the first stage, the LP relaxation model is solved; in the second stage,  some techniques are applied to determine the values of elements of the LP solution that do not satisfy the one-bit constraint. Different techniques in the second stage lead to different algorithms.
In particular, the partial branch-and-bound (P-BB) algorithm and the ordered partial sequential update (OPSU) algorithm proposed in \cite{PBB} apply a BB procedure and a greedy procedure in the second stage, respectively. The CI-based approaches generally enjoy significantly better performance than the MMSE-based approaches. However, their performance degrades in large-scale systems with high-order modulation (e.g., OPSU) or their computational costs are prohibitively high (e.g., P-BB). 

In this paper, we focus on the CI-based symbol scaling model for one-bit downlink transmission with phase shift keying (PSK)  modulation. We propose an efficient negative $\ell_1$ penalty (NL1P) approach for solving the considered problem, which is especially efficient in the massive MIMO scenario where the problem dimension is large. More specifically, we first introduce a novel negative $\ell_1$ penalty model, which shares the same global and local solutions with the original problem when the penalty parameter is sufficiently large. This is in sharp contrast to the LP relaxation model on which the existing approaches (e.g., MSM, OPSU, P-BB) are based. Then, we transform the penalty model  into an equivalent min-max problem. By taking care of its special structure, we propose an efficient alternating optimization (AO) algorithm for solving the reformulated min-max problem, where at each iteration only two matrix-vector multiplications and one projection onto the simplex need to be computed, making it particularly suitable for solving large-scale problems. We also establish the global convergence of the AO algorithm.  Simulation results show that our proposed algorithm achieves a better tradeoff between the bit-error-rate (BER) performance and the computational efficiency than the state-of-the-art CI-based algorithms.
\vspace{-0.2cm}
\section{problem formulation}
\subsection{System Model}
Consider a downlink multiuser massive MIMO system in which a BS equipped with $N_t$  antennas transmits signals to $K$ single-antenna users simultaneously. The received signal vector $\by\in\mathbb{R}^{K\times 1}$ is given by\vspace{-0.2cm}
$$\by=\bH\bx_T+\bn,$$
where $\bH=[\bh_1,\dots,\bh_K]^\T\in \mathbb{C}^{K\times N_t}$ is the flat-fading channel matrix between the BS and the users, $\bx_T$ is the transmitted signal, and $\bn\sim\mathcal{C}\mathcal{N}(\mathbf{0},\sigma^2\mathbf{I})$ is the additive white Gaussian noise. 

 We consider the scenario where one-bit DACs are employed at the BS. In this case,  each element of $\x_T$ is drawn from a discrete set consisting of only four symbols. In particular, $\bx_T\in\left\{\pm\frac{1}{\sqrt{2N_t}}\pm\frac{1}{\sqrt{2N_t}}j\right\}^{N_t}$, where $j$ is the imaginary unit 
 (satisfying $j^2=-1$)
 and $\x_T$ is normalized to be of unit norm. In this paper, we restrict our attention to PSK modulation, that is, all elements of the intended data symbol vector $\bs=[s_1,\dots, s_K]^\T$ for the users are drawn from a unit-norm $M$-PSK modulation. 
Our goal here is to design the transmitted signal $\x_T$ such that the SEP is as low as possible. 
\vspace{-0.1cm}
\subsection{Problem Formulation}
We adopt the CI-based symbol scaling model to formulate our interested problem as in \cite{CImodel, PBB}. The main idea is to maximize the minimum distance from all received noise-free signals to their corresponding decision boundaries.
\begin{figure}
\centering
\includegraphics[scale=0.24]{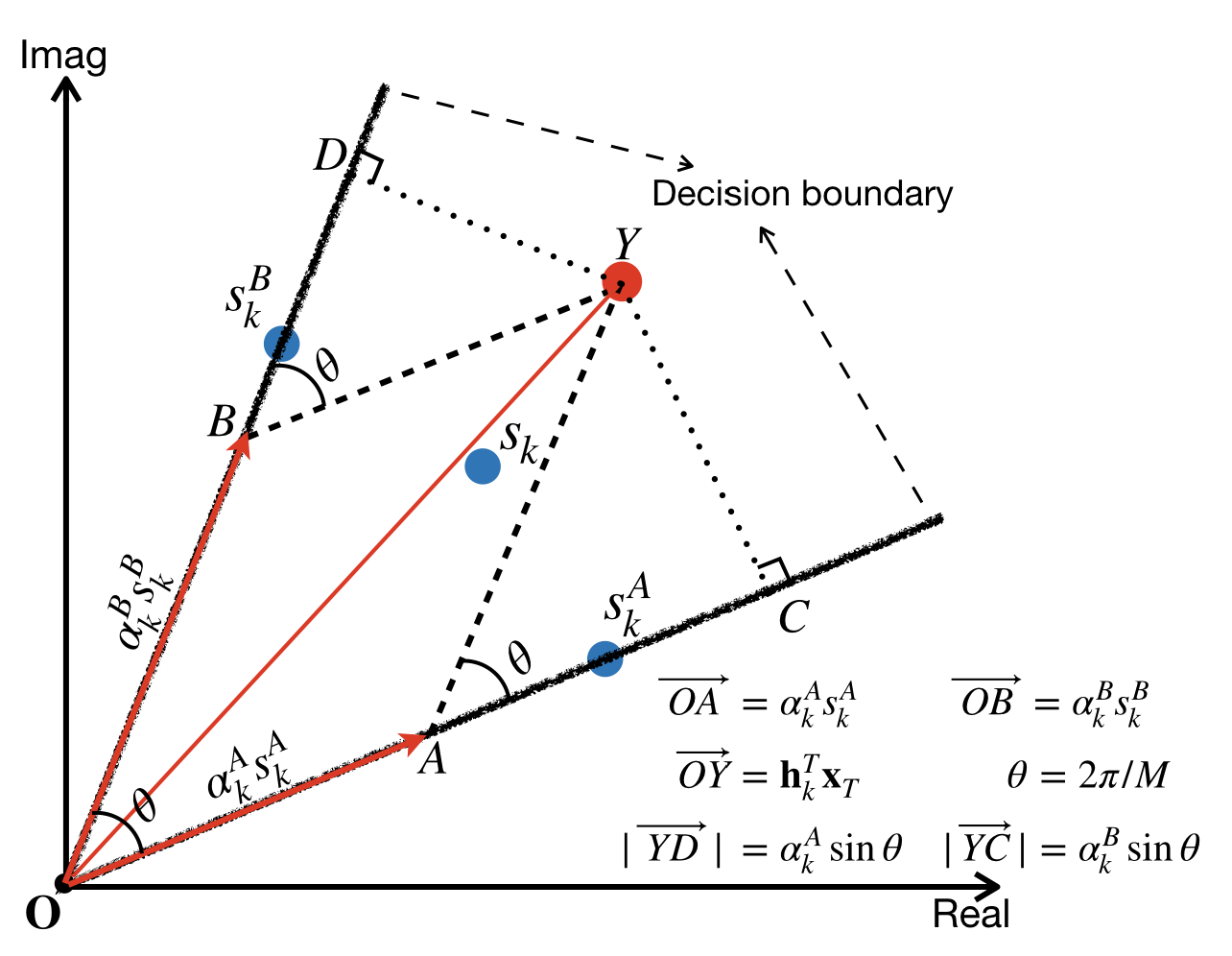}
\vspace{-0.25cm}
\caption{An illustration of the CI formulation for 8-PSK.}
\label{fig3}
\end{figure}
 Taking $8$-PSK modulation as an example and assuming the intended data symbol for user $k$ is $s_k=e^{j\pi/4}$, we illustrate in Fig. \ref{fig3} how to characterize the distance from the noise-free received signal $\hat{y}_k$ (corresponding to $\overrightarrow{OY}$) to its decision boundary.  In particular, we decompose $\hat{y}_k$ along $s_k^A$ and $s_k^B$, which are the unit vectors parallel to the two decision boundaries of $s_k$,  as 
$$\hat{y}_k=\alpha_k^As_k^A+\alpha_k^Bs_k^B.$$
Note that $\theta=\frac{2\pi}{M}$ is a constant when the constellation level $M$ is given, and thus the distance $\min\left\{|\overrightarrow{YD}|,|\overrightarrow{YC}|\right\}$ is only determined by $\min\left\{\alpha_k^A,\alpha_k^B\right\}$, which leads to the following model for CI-based one-bit precoding design \cite{CImodel, PBB}:\vspace{-0.081cm}
\begin{subequations}
\begin{align}
\max_{\bx_T}~&\min_{k\in\{1,2,\dots,K\}}~\left\{\alpha_k^A,\alpha_k^B\right\}\notag\\
\hspace{-0.2cm}\text{(}\text{P}_0\text{)}\hspace{0.4cm}\text{s.t. }~&\bh_k^\T\bx_T=\alpha_k^As_k^A+\alpha_k^Bs_k^B,\quad k=1, 2,\dots,K,\label{2a}\\
~&\bx_T(i)\in\left\{\pm1 \pm j\right\}, \quad i=1, 2,\dots, N_t,
\end{align}
\end{subequations} 
where
we remove the problem-dependent quantity $\frac{1}{\sqrt{2N_t}}$ from the constraint on $\x_T$ and incorporate it into $\bH$. Since $\alpha_k^A$ and $\alpha_k^B$ are both real numbers, we can express $[\alpha_k^A, \alpha_k^B]^\T$ explicitly as a function of $\bh_k$, $s_k$, and $\bx=[\RR(\bx_T)^\T,\I(\bx_T)^\T]^\T$ by rewriting the complex-valued constraints (\ref{2a}) into the real-valued form. Moreover, the original maximization problem can be converted into a minimization problem (by adding a negative sign in the objective). Then we arrive at the following compact form: 
\begin{equation}\label{maxminform}
\begin{aligned}
\min_{\bx}~&\max_{l\in\{1,2,\dots,2K\}}\alpha_l\\
\qquad\text{s.t. }~&\boldsymbol{\Lambda}=\mathbf{A}\bx,\\
~&\bx(i)\in\left\{-1,1\right\},~i=1,2,\dots, 2N_t,
\end{aligned}
\end{equation} where $\boldsymbol{\Lambda}=-
\left[\alpha_1^A, \alpha_1^B, \alpha_2^A, \alpha_2^B, \dots, \alpha_K^A, \alpha_K^B\right]^\T\triangleq[\alpha_1,\alpha_2,\dots,\alpha_{2K}]^\T\hspace{-0.13cm}\in\hspace{-0.05cm}\mathbb{R}^{2K}$ \hspace{-0.07cm}and $\bA=-\left[\mathbf{V}_1^\T,\mathbf{V}_2^\T,\dots,\mathbf{V}_K^\T\right]^\T\in\mathbb{R}^{2K\times 2N_t}$ with

$$\mathbf{V}_k=
\displaystyle\frac{
\left[
\begin{matrix}
\I(s_k^B)&-\RR(s_k^B)\\-\I(s_k^A)&\RR(s_k^A)
\end{matrix}\right]
\left[
\begin{matrix}
\RR(\bh_k^\T)&-\I(\bh_k^\T)\\\I(\bh_k^\T)&\RR(\bh_k^\T)
\end{matrix}\right]}{\RR(s_k^A)\I(s_k^B)-\I(s_k^A)\RR(s_k^B)}.$$
See \cite{CImodel,PBB} for detailed derivations.

The constraint $\boldsymbol{\Lambda}=\mathbf{A}\bx$ in problem \eqref{maxminform} can be further substituted into the objective, which leads to the following  form: 
\begin{equation*}\label{Pe}
\text{(P)}\qquad\min_{\x\in\{-1,1\}^{n}}\max_{l\in\{1,2,\dots,m\}}\mathbf{a}_l^\T\x,
\end{equation*}
where $n=2N_t,$ $m=2K,$ and $\mathbf{a}_l^\T$ is the $l$-th row of $\mathbf{A}$. 
In the following, we shall design algorithms based on the compact form (P), which appears to be easier to handle than the form (P$_0$).
\section{Proposed Negative $\ell_1$ Penalty Approach}

Solving problems with a non-smooth objective and discrete constraints like (P) is generally challenging. In addition, the considered massive  MIMO scenario leads to large-scale problems, which places high demand on the efficiency of the algorithm. In this section, we propose an efficient negative $\ell_1$ penalty approach for finding a high-quality solution of problem (P).

\subsection{Exact Penalty Model for Problem (P)} To deal with the discrete one-bit constraint in (P), we resort to the penalty technique. More specifically, we transform problem (P) into the following negative $\ell_1$ penalty model\footnote{A closely related work \cite{GEMM} considers to first smooth the objective in (P) and then apply the negative square penalty, i.e., $-\|\bx\|_2^2$, to the smoothed problem.  In contrast, our proposed penalty model deals with the original non-smooth objective, in which case the exact penalty property (see Theorem \ref{theorem3} further ahead) does not hold for the negative square penalty, and thus the non-smooth negative $\ell_1$ penalty is adopted in this paper.}:
\begin{equation*}
\text{(}\text{P}_\lambda\text{)} \qquad\min_{\x\in[-1,1]^{n}}\max_{l\in\{1,2,\dots,m\}}~\mathbf{a}_l^\T\x-\lambda\|\x\|_1,
\end{equation*}
in which the discrete constraint is relaxed and a negative $\ell_1$-norm term is included in the objective to encourage large magnitude of $\{x_i\}$. We establish the equivalence between the penalty problem (P$_\lambda$) and the original problem (P) in the following theorem, whose proof can be found in \cite{journal} and \cite{tech}. 
\begin{dingli}[Exactness of Penalty Model (P$_\lambda$)]\label{theorem3}
If the penalty parameter $\lambda$ in (P$_{\lambda}$) satisfies $\lambda>\max_l\|\mathbf{a}_l\|_\infty$, then the following results hold:
\begin{enumerate}
\renewcommand{\labelenumi}{(\theenumi)}
\item Any optimal solution of (P$_\lambda$) is also an optimal solution of (P), and vice versa.
\item Any local minimizer of (P$_\lambda$) is a feasible point of (P); on the other hand, any feasible point of (P) is also a local minimizer of (P$_\lambda$).
\end{enumerate}
\end{dingli}
The above theorem reveals that problem (P$_\lambda$) is an exact reformulation of problem (P) in the sense that the two problems share the same global and local solutions. This motivates us to solve the discrete  problem (P) by solving the continuous problem (P$_\lambda$).

\subsection{Min-Max Reformulation of the Penalty Model} Problem (P$_\lambda$) is still challenging to solve due to its non-smooth and non-convex objective. To tackle it, we introduce an auxiliary variable 
$\by\in\Delta\triangleq\left\{\by\in\mathbb{R}^m\mid\mathbf{1}^\T\by=1, \by\geq\mathbf{0}\right\}$,  where $\mathbf{1}$ denotes the all-one vector,
to reformulate problem (P$_\lambda$) as the following min-max problem:
\begin{equation*}
\text{(}\widehat{\text{P}}_\lambda\text{)}\quad\min_{\bx\in[-1,1]^n}~{\max_{\by\in\Delta}~\by^\T\bA\bx}-\lambda\|\bx\|_1.
\end{equation*}
 It is shown in \cite{HiBSA} that the two problems (P$_\lambda$) and ($\widehat{\text{P}}_\lambda$) are equivalent. In particular,  an optimal solution (stationary point) of one problem can be easily constructed given an optimal solution  (stationary point) of the other problem. In the following, we design an efficient algorithm for problem ($\widehat{\text{P}
}_\lambda$) by exploiting its special structure. 

\subsection{Alternating Optimization Algorithm for ($\widehat{\text{P}}_\lambda$)} Note that if the variable $\by$ in problem ($\widehat{\text{P}}_\lambda$) is fixed, then the objective is separable in $\bx$. Based on this, we consider to update $\bx$ and $\by$ in an alternating fashion.
 
Our proposed algorithm can be regarded as an extension of the algorithms proposed in \cite{HiBSA} and \cite{xu2020unified}, which are designed for smooth min-max problems and thus cannot be applied directly to our interested problem ($\widehat{\text{P}}_\lambda$). Similar to \cite{HiBSA} and \cite{xu2020unified},  we consider a perturbed function:\vspace{-0.28cm}
$$f(\bx,\by)=\by^\T\bA\x-\lambda\|\bx\|_1-\frac{c_k}{2}\|\by\|_2^2,\vspace{-0.2cm}$$
where the perturbed term is introduced to make $f(\bx,\by)$ strongly concave in $\by$. It is shown in \cite{HiBSA} and \cite{xu2020unified} that the perturbed term is important for the convergence of the corresponding algorithms.
At each iteration, our proposed algorithm performs the following updates:\vspace{-0.2cm}
\begin{subequations}
\begin{align}
\bx_{k+1}&\in\arg\min_{\bx\in[-1,1]^n}\by_k^\T\bA\bx-\lambda\|\bx\|_1+\frac{\tau_k}{2}\|\bx-\bx_k\|_2^2,\label{updatex}\\
\by_{k+1}&=\text{Proj}_{\Delta}\left(\by_k+{\rho_k}\bA\bx_{k+1}-{\rho_k}c_k\by_k\right),\label{updatey}
\end{align}
\end{subequations}
where $\rho_k\geq 0, \tau_k\geq 0,$ and $c_k\geq 0$ are the parameters that need to be selected carefully (and the choices of these parameters will be specified later in Theorem \ref{The2}).  Since the above algorithm updates $\bx$ and $\by$ alternately, we name it as the alternating optimization (AO) algorithm.

The update of variable $\bx$ is a normal step which minimizes the current objective plus a regularization term. 
It is easy to check that the $\bx$-subproblem \eqref{updatex} admits a closed-form solution as
\vspace{-0.2cm}
\begin{equation}\label{solutionx}
\x_{k+1}(i)=\text{sgn}(a_k^i)\min\left\{|a_k^i|+\frac{\lambda}{\tau_k},1\right\},~i=1,2,\dots,n, 
\vspace{-0.2cm}
\end{equation}
where $a_k^i=\bx_{k}(i)-\frac{\bA_i^\T\by_k}{\tau_k}$, $\bA_i$ denotes the $i$-th column of $\bA$, and $\text{sgn}(\cdot)$ returns the sign of the corresponding real number.
The update of variable $\by$ is a projection gradient step for the perturbed function.
 The solution of the $\by$-subproblem \eqref{updatey} involves only one matrix-vector multiplication  and one projection onto the simplex, which has a very fast implementation \cite{projection}. Therefore, the proposed AO algorithm enables us to solve problem ($\widehat{\text{P}}_\lambda$) very efficiently. We summarize the AO algorithm for solving problem ($\widehat{\text{P}}_\lambda$) in Algorithm \ref{algonp}.
 
\vspace{-0.1cm}
\begin{algorithm}
\caption{Proposed AO Algorithm for Solving Problem ($\widehat{\text{P}}_\lambda$)}
\begin{algorithmic}\label{algonp}
\small
\STATE Step 1 \ Input $\bx_0,\by_0,\{\tau_k\},\{\rho_k\}, \{c_k\}$; set $k=1$.
\STATE Step 2 \ Alternately update $\bx_k$ and $\by_k$ as in \eqref{solutionx} and \eqref{updatey}.
\STATE Step 3 \ If some stopping criterion is satisfied, stop; otherwise, set $k=k+1$, go to Step 2.
\end{algorithmic}
\end{algorithm}

\vspace{-0.2cm}
Next we present the convergence results of the proposed AO Algorithm (and its proof is provided in \cite{tech}).  In particular, the following theorem shows that when the penalty parameter $\lambda$ is sufficiently large and the algorithm parameters are properly selected, every limit point $\hat{\bx}$ of the sequence generated by Algorithm \ref{algonp} is a local minimizer of problem (P$_\lambda$), and more importantly, it satisfies the one-bit constraint. This desirable property is a combination of nice properties of the penalty model (P$_\lambda$) and Algorithm \ref{algonp}.
\begin{dingli}\label{The2}
Let $\{(\bx_k,\by_k)\}$ be the sequence generated by Algorithm \ref{algonp} with $\rho_k=\rho, c_k=\frac{\beta_1}{ k^{\gamma}}$, and $\tau_k=\frac{16\beta_2\|\bA\|_2^2}{\rho c_k^2}+\beta_3$, where $0<\rho\leq\frac{1}{\beta_1}$, $0<\gamma\leq 0.5$, $\beta_1>0$, $\beta_2>1$, and $\beta_3\geq\rho\|\bA\|_2^2$. Then every limit point $(\hat{\bx},\hat{\by})$ of $\{(\bx_k,\by_k)\}$ is a stationary point of problem ($\widehat{\text{P}}_\lambda$). Moreover, if $\lambda>\max_l\|\ba_l\|_\infty,$ $\hat{\bx}$ is  a local minimizer of problem (P$_\lambda$) and satisfies the one-bit constraint.
\end{dingli}
\subsection{Negative $\ell_1$ Penalty Approach for Problem (P)}
Theorems \ref{theorem3} and \ref{The2} inspire us to find a high-quality solution of problem (P) by applying Algorithm \ref{algonp} to solve problem ($\widehat{\text{P}}_\lambda$) (equivalent to problem (P$_\lambda$))  with a sufficiently large penalty parameter $\lambda$. To  further improve the numerical performance, we employ the homotopy/continuation technique \cite{homotopy1,penaltyproof}, i.e., we initialize the penalty parameter with a small value at the beginning, then gradually increase it and trace the solution path of the corresponding penalty problems, until the penalty parameter is sufficiently large and a one-bit solution is obtained.  We name the whole procedure for solving problem (P) as the negative $\ell_1$ penalty (NL1P) approach and summarize it as follows.
\vspace{-0.1cm}
\begin{algorithm}[H]
\caption{Proposed NL1P Approach for Solving Problem (P)}
\label{nl1p}
\begin{algorithmic}
\small
\STATE Step~\hspace{0.015cm}1 ~Input $\lambda^{(0)},\delta>1,$ $\bx^{(0)}$; set $t=1$.
\STATE Step~2 ~\hspace{-0.1cm}Apply Algorithm \ref{algonp} to solve problem (${\text{P}}_{\lambda}$) with parameter $\lambda=\lambda^{(t-1)}$ and initial point $\bx^{(t-1)}$; let the solution be $\x^{(t)}$.

\STATE Step~3 ~\hspace{-0.1cm}Stop if $\bx^{(t)}$ satisfies the one-bit constraint; otherwise, set $\lambda^{(t)}=\delta\lambda^{(t-1)}$ and $t=t+1$, go to Step 2.
\end{algorithmic}
\end{algorithm}

\vspace{-0.5cm}
\section{Simulation Results}
In this section, we present simulation results to show both the effectiveness and the efficiency of our proposed NL1P approach. We consider multiuser massive MIMO systems where the BS is equipped with hundreds of antennas. The transmission block length is set to be $L=10$ and the SNR is defined as $\frac{1}{\sigma^2}$, where the unit transmit power is assumed. The channel matrix $\bH$ is composed of independent and identically distributed Gaussian random variables with zero mean and unit variance. All the results are obtained with Monte Carlo simulations of 1000 independent channel realizations. 

The parameters used in our algorithms are as follows. In Algorithm \ref{nl1p}, the initial point is chosen as  $\x^{(0)}=\mathbf{0}$; the penalty parameter is initialized as $\lambda^{(0)}=\frac{0.001M}{8}$ and increased by a factor of $\delta=5$ at each iteration. In Algorithm \ref{algonp}, we set the initial point of $\by$ as $\by_0=\frac{1}{2K}\mathbf{1}$, and the other parameters  as $  \rho_k=\rho=\frac{0.2}{\|\bA\|_2},~c_k=\frac{0.01}{\rho k^{0.05}},$ and $ \tau_k=\frac{2\log_2N_t+1}{10}\text{mean}\left(|\bA|\right)k^{0.1}$. We terminate Algorithm \ref{algonp} for solving the subproblem (${\text{P}}_\lambda$) in Algorithm \ref{nl1p} when its iteration number is more than $500$ or when the distance of its successive iterates is less than $10^{-3}$.

We compare the proposed NL1P approach with the following algorithms: zero-forcing (ZF) with infinite-resolution DACs, termed as `Inf-Bit ZF', which serves as the performance limit of all one-bit precoders; ZF followed by one-bit quantization \cite{linear1}, termed as `1-Bit ZF';  SQUID \cite{SQUID} which is an algorithm based on the MMSE metric, termed as `MMSE 1-Bit SQUID'; the MSM precoder \cite{CIfirst} based on the CI metric obtained by quantizing the LP relaxation solution, termed as `CI 1-Bit MSM';  OPSU and P-BB \cite{PBB} based on the CI metric, termed as `CI 1-Bit OPSU' and `CI 1-Bit P-BB', respectively.

\begin{figure}[t]
\centering
\includegraphics[scale=0.334]{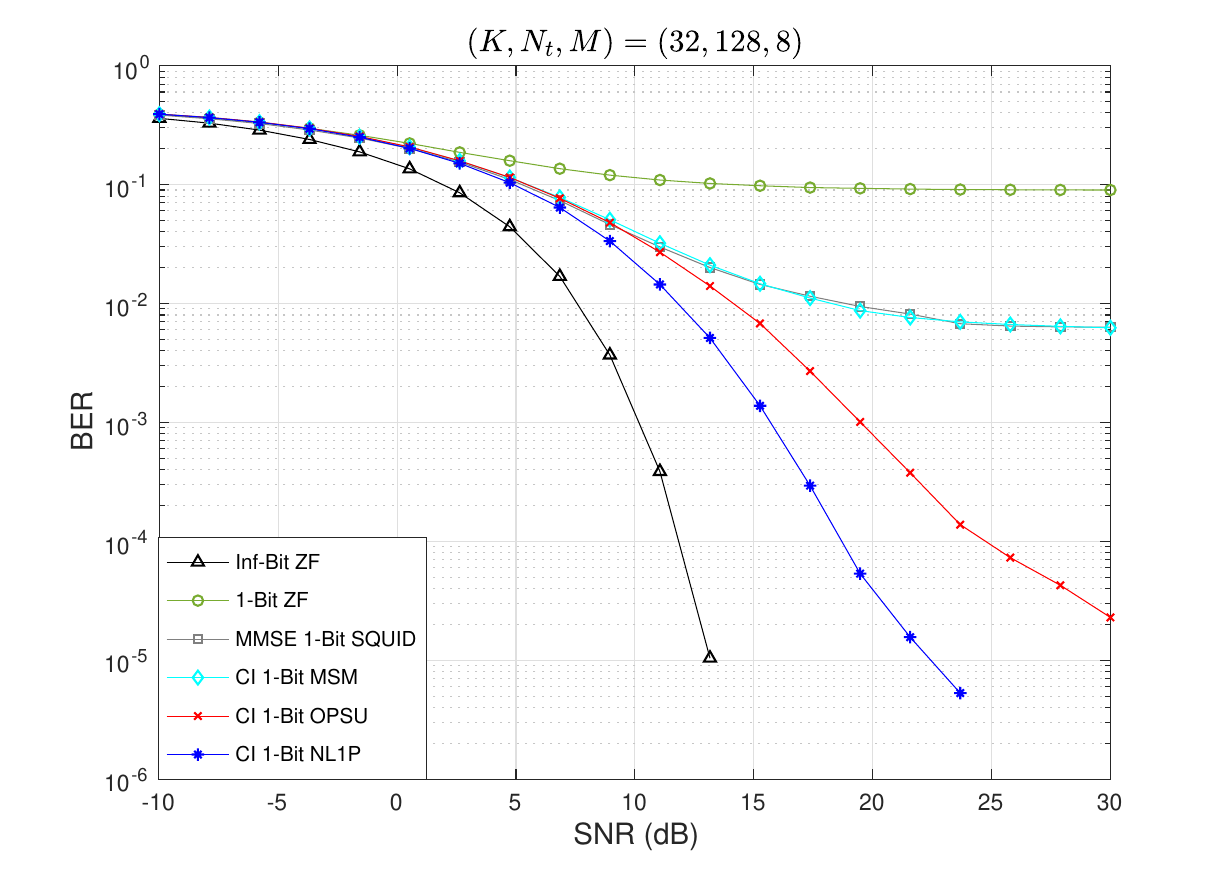}
\vspace{-10pt}
\caption{BER performance versus SNR, where $(K,N_t,M)=(32,128,8)$.}
\label{32_128_8}
\end{figure}
\begin{figure}[t]
\centering
\includegraphics[scale=0.334]{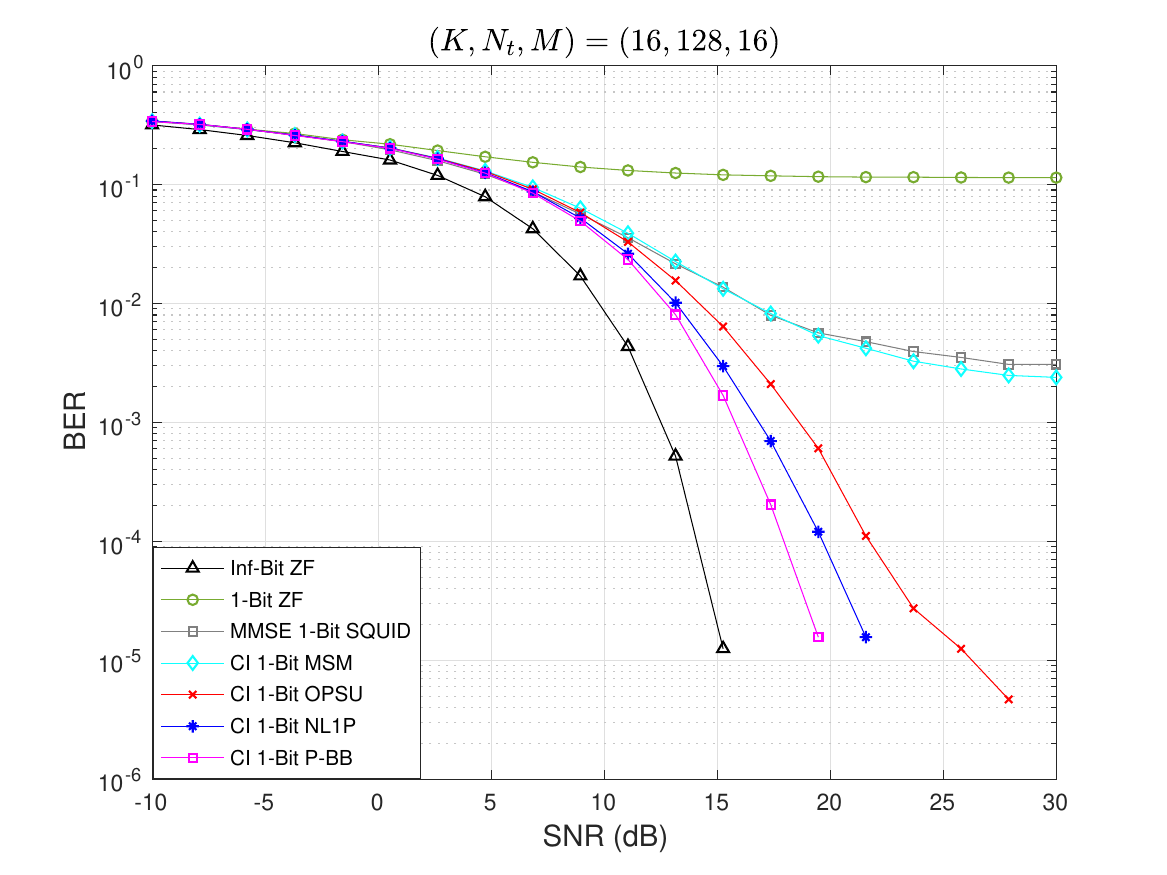}
\vspace{-10pt}
\caption{BER performance versus SNR, where $(K,N_t,M)=(16,128,16)$.}
\label{16_128_16}
\end{figure}
\begin{figure}[t]
\centering
\includegraphics[scale=0.334]{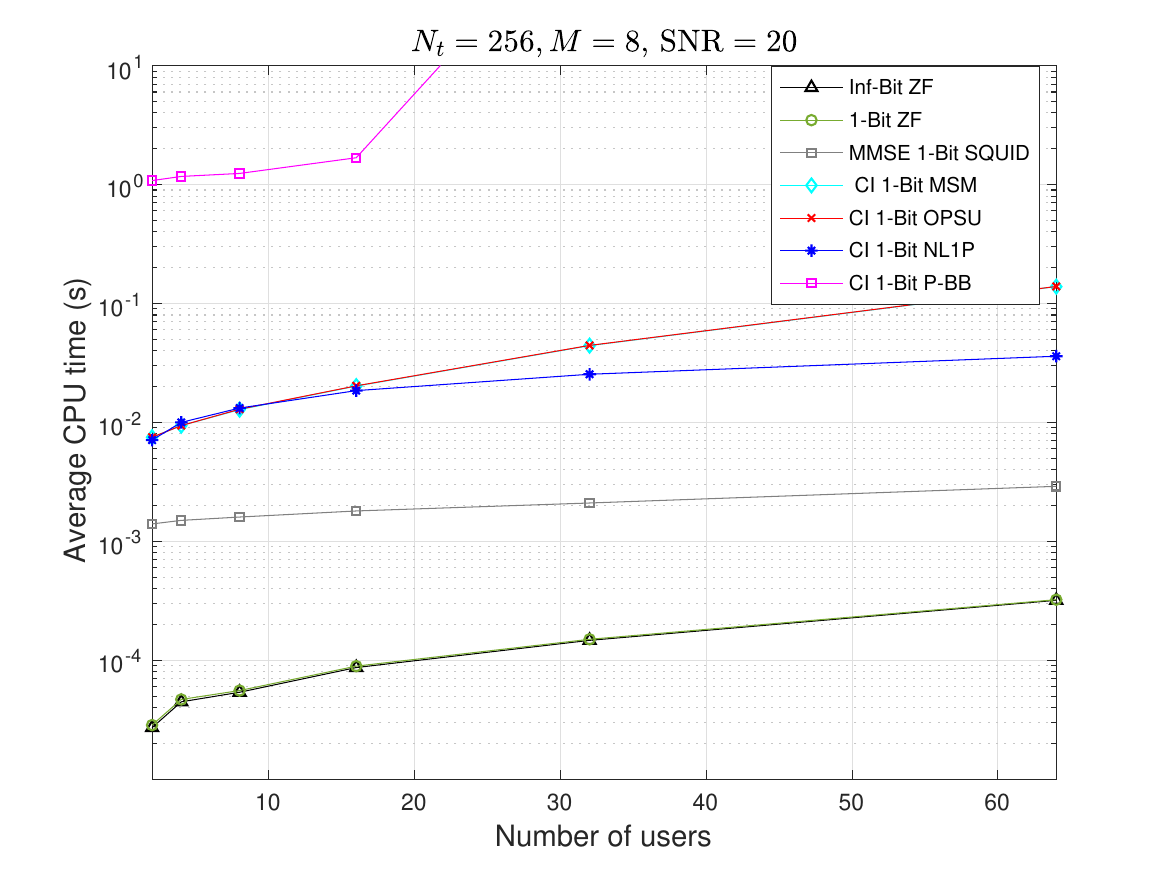}
\vspace{-10pt}
\caption{CPU time versus the number of users, where $N_t=256$, $M=8$, and SNR$=20$.}
\label{time_user}
\end{figure}

In Figs. \ref{32_128_8} and  \ref{16_128_16}, we present the BER results for different massive MIMO systems. Specifically, in Fig. \ref{32_128_8} we consider a $32\times 128$ system with $8$-PSK modulation and in Fig. \ref{16_128_16} we consider a $16\times 128$ system with $16$-PSK modulation. The P-BB approach is not included in Fig. \ref{32_128_8} due to its prohibitively high complexity. As shown in the figures, the one-bit ZF precoder suffers a severe BER floor due to its coarse one-bit quantization, while all of the nonlinear approaches exhibit significantly better BER performance. Nevertheless, the MMSE-based SQUID approach and the CI-based MSM approach also saturate early in the high SNR regime.  Of the precoders that offer satisfactory performance, the proposed approach exhibits better error-rate performance than the state-of-the-art OPSU precoder. In particular, we can observe an SNR gain up to nearly $6$dB and $2.5$dB  in Fig. \ref{32_128_8} and Fig. \ref{16_128_16} respectively when the BER is $10^{-4}$; as the BER becomes lower, the performance gain in terms of the SNR also becomes larger. The P-BB algorithm, though with slightly better performance than our proposed algorithm, is much more computationally inefficient, as will be demonstrated in Fig. \ref{time_user}.

In Fig. \hspace{-0.06cm}\ref{time_user}, we evaluate the efficiency of the compared algorithms by reporting their CPU time.  Among all the compared CI-based precoders, our proposed approach is the most efficient. More specifically, the computational costs of the MSM precoder and the OPSU precoder increase rapidly with the scale of the system, while that of our proposed  approach grows much slower. This is because both of the MSM and OPSU  algorithms solve the LP relaxation model via the interior-point method, whose complexity is high when the problem dimension is large, while the proposed NL1P approach solves the penalty model (P$_\lambda)$ with the AO algorithm, which enjoys low per-iteration complexity. As shown in the figure, the P-BB algorithm is much more computationally expensive than all the other methods. Its computational cost becomes prohibitively high when the number of users is large, since the complexity of the branch and bound procedure grows exponentially with respect to the number of users \cite{PBB}.  This makes the P-BB approach unsuitable for practical implementation and can only serve as a performance benchmark.

From the simulation results, we can conclude that our proposed NL1P approach achieves a better tradeoff between the BER performance and the computational efficiency than the state-of-the-art CI-based algorithms. The good BER performance  is mainly attributed to the exactness of the negative $\ell_1$ penalty model and the high computational efficiency is due to the efficiency of the AO algorithm for solving the penalty model (P$_\lambda$) in the proposed NL1P approach.
\bibliographystyle{IEEEtran}
\bibliography{reference}

\end{document}